\documentclass{article}

\usepackage{spconf}

\usepackage{
    amsfonts,
    amsmath,
    color,
    colortbl,
    commath,
    graphicx,
    hyperref,
    soul,
    wasysym
}

\definecolor{tablecolor}{cmyk}{0,0,0,0.12}

\title{Crowdsourced and Automatic Speech Prominence Estimation}
\name{Max Morrison, Pranav Pawar, Nathan Pruyne, Jennifer Cole, Bryan Pardo} 
\address{Northwestern University, Evanston, IL, USA}

\begin{document}
\ninept
\maketitle


\begin{abstract}
The prominence of a spoken word is the degree to which an average native listener perceives the word as salient or emphasized relative to its context. Speech prominence estimation is the process of assigning a numeric value to the prominence of each word in an utterance. These prominence labels are useful for linguistic analysis, as well as training automated systems to perform emphasis-controlled text-to-speech or emotion recognition. Manually annotating prominence is time-consuming and expensive, which motivates the development of automated methods for speech prominence estimation. However, developing such an automated system using machine-learning methods requires human-annotated training data. Using our system for acquiring such human annotations, we collect and open-source crowdsourced annotations of a portion of the LibriTTS dataset. We use these annotations as ground truth to train a neural speech prominence estimator that generalizes to unseen speakers, datasets, and speaking styles. We investigate design decisions for neural prominence estimation as well as how neural prominence estimation improves as a function of two key factors of annotation cost: dataset size and the number of annotations per utterance.
\end{abstract}




\section{Introduction}
\label{sec:intro}

\textit{Prominent} or \textit{emphasized} words are those that stand out to listeners as salient or perceptually highlighted relative to their context. Human perception of prominence in English (and many other languages) is influenced by acoustic factors related to \textit{prosody} (i.e., the pitch, rhythm, and loudness of speech) as well as \textit{information structure}~\cite{COLE2019113, 0627d41479f8467dbc1d64af09aa0e85}. Information structure is the contribution a word makes to the shared knowledge of the speaker and hearer, based on its status as conveying information that is given, new, or contrastive relative to prior discourse. Prominence is a multi-dimensional perceptual relation between words in a phrase. However, performing annotations in this multi-dimensional space (e.g., separately annotating structural and semantic factors), as opposed to a scalar, is cost-prohibitive. 

In this paper, we represent the emphasis status of each word using a binary label (zero or one), and its prominence as a scalar real value between zero and one, such that the emphasis status is a binary thresholding of scalar prominence, and scalar prominence is a weighted norm of the latent multi-dimensional prominence. From a statistical viewpoint, we consider the prominence $m$ of word $i$ as a Bernoulli distribution parameter and the emphasis $e$ as the corresponding Bernoulli random variable such that the probability that word $i$ is perceived as emphasized is $p(e_i = 1) = m_i$.

Emphasis annotation is the process of labeling each word with a binary indicator of emphasis, a task that can be performed by non-expert native speakers. Given multiple annotations of the same speech (e.g., from multiple human annotators), scalar prominence values can be obtained by averaging over the binary emphasis annotations for each word. Emphasis and prominence labels are used in downstream tasks such as emphasis-controlled TTS~\cite{seshadri22_interspeech, Suni_2020, roekhaut2010model}, emotion recognition~\cite{rao_speech_tech_2013}, and text summarization~\cite{Chen_Pan_2017}.

Because human emphasis annotation can be costly and time-consuming, prior works attempted to replace these annotations with one of three types of automated methods: (1) heuristic, rule-based methods based on acoustic features~\cite{KAKOUROS201667, SUNI2017123, 9383591}, (2) machine learning methods that train on the output of such a rule-based system~\cite{malisz17_interspeech, talman2019predicting, stephenson22_interspeech}, and (3) machine learning methods that train on ground-truth values prominence derived from human annotation~\cite{mishra12_interspeech, Christodoulides2017, stehwien2020acoustic, 9747780}. Heuristic, rule-based systems struggle to capture the complexity of high-dimensional perceptual attributes and can require significant manual tuning to generalize to new data distributions. Machine learning methods that learn from the output of rule-based systems might perform useful interpolation and denoising, but otherwise inherit the same drawbacks. Methods that learn from human annotation avoid these drawbacks, but require human-annotated data for training. All three approaches necessitate benchmarking on human annotations, making human annotation unavoidable.

Vaidya et al.~\cite{9747780} is most similar to our method of performing neural prominence estimation using human annotations. They use a closed-source dataset of 41k English words spoken by 10-14 year old students in Mumbai and annotated by 26 university students to have seven annotations per word. Transcripts come from 34 short stories of ~100 words, each selected to have a single reference prosody for fluency assessment purposes~\cite{SABU2021101200}. In other words, text was selected to make placement of emphasis in the dataset more predictable. Vaidya et al. train a CRNN-based model to infer prominence from ground truth acoustic features with a Pearson correlation of 0.721 on heldout recordings of unseen non-native child speakers. They demonstrate further performance using human-annotated boundary features, language-dependent lexical features, and curated selection of prosodic features. Lack of open source data, models, and input features makes exact replication impossible. Instead, we perform ablations of what we consider to be key architectural decisions and note the additional utility of the input features they have explored.

Unique to our work, we train on scalable crowdsourced annotations, generalize to unseen speaking styles and datasets of adult speakers, produce open-source annotations suitable for training and benchmarking, develop open-source tools for crowdsourced speech annotation, and uncover novel guidelines for how the performance of neural prominence estimation trained on crowdsourced annotations scales with dataset size and annotator redundancy---two primary factors in the overall cost of emphasis annotation. Our main contributions in this paper are as follows.


\begin{figure*}[ht]
    \centering
    \includegraphics[width=\linewidth]{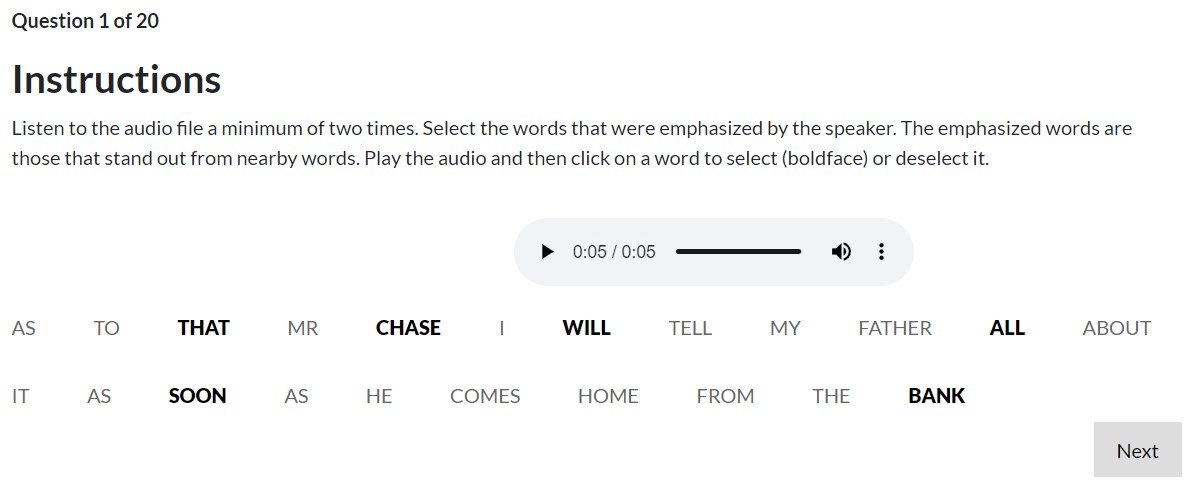}
    \vspace{-2em}
    \caption{\textbf{Crowdsourced human emphasis annotation interface $|$} Annotators listen to a speech recording and click to boldface the words they perceive as emphasized.}
    \label{fig:word-selection}
    \vspace{-1em}
\end{figure*}

\begin{itemize}
    \item \textbf{(Contribution 1)} We develop a neural speech prominence estimation system trained on crowdsourced human emphasis annotations that produces accurate prominence estimations on unseen speakers, datasets, and speaking styles (Section~\ref{sec:model}).
    \item \textbf{(Contribution 2)} We produce a CC-BY-4.0 licensed dataset of emphasis annotations of one eighth of the \\\texttt{train-clean-100} partition of LibriTTS (Section~\ref{sec:dataset}).
    \item \textbf{(Contribution 3)} We develop an open-source system for performing crowdsourced word-level annotations of, e.g., falsetto, vocal fry, and emphasis (Section~\ref{sec:annotation}). 
    \item \textbf{(Contribution 4)} We demonstrate how the amount of training data and the number of annotators per speech excerpt impact estimation performance, providing guidelines for cost-effective annotation (Section~\ref{sec:results}).
\end{itemize}

\noindent
We release our code and annotation methods as \texttt{emphases}, a MIT-licensed, pip-installable Python module for training, evaluating, and performing both automatic and human annotation of emphasis. Our code and dataset are available on our project website.\footnote{\texttt{\href{https://www.maxrmorrison.com/sites/prominence-estimation/}{maxrmorrison.com/sites/prominence-estimation/}}}


\section{Crowdsourcing emphasis annotation}
\label{sec:annotation}

Here, we describe our open-source crowdsourced annotation tool for annotation of, e.g., prominence, mispronunciation, or vocal fry annotation. We developed our human annotation system as a word selection task that we add to Reproducible Subjective Evaluation (ReSEval)~\cite{morrison2022reproducible}. ReSEval is a subjective evaluation tool that handles database, server, and crowdsourced participant acquisition to quickly create and manage crowdsourced evaluations in Python. ReSEval enables a greater variety of tasks and prescreening criteria (e.g., listening tests) than existing survey templates, such as those available on Amazon Mechanical Turk (MTurk). For prominence annotation, we first require annotators to pass a listening test that ensures a suitable listening environment, using the listening test method proposed by Cartwright et al.~\cite{cartwright2016fast}. We then present annotators with an audio recording and the corresponding text (see Figure~\ref{fig:word-selection}). As in the annotation interface of Cole et al.~\cite{cole2017crowd}, annotators are required to listen to the audio file a minimum of two times and asked to select all of the words that were emphasized by the speaker by clicking on the words themselves. Annotators must start the audio to begin selecting words, but may select words while the audio is playing.


\section{Emphasis annotation dataset}
\label{sec:dataset}

We used our crowdsourced annotation system to perform human annotation on one eighth of the \texttt{train-clean-100} partition of the LibriTTS~\cite{zen2019libritts} dataset. Specifically, participants annotated 3,626 utterances with a total length of 6.42 hours and 69,809 words from 18 speakers (9 male and 9 female). We collected at least one annotation of all 3,626 utterances, at least two annotations of 2,259 of those utterances, at least four annotations of 974 utterances, and at least eight annotations of 453 utterances. We did this in order to explore (in Section~\ref{sec:results}) whether it is more cost-effective to train a system on multiple annotations of fewer utterances or fewer annotations of more utterances. We paid 298 annotators to annotate batches of 20 utterances, where each batch takes approximately 15 minutes. We paid \$3.34 for each completed batch (estimated \$13.35 per hour). Annotators each annotated between one and six batches. We recruited on MTurk US residents with an approval rating of at least 99 and at least 1000 approved tasks. Today, microlabor platforms like MTurk are plagued by automated task-completion software agents (bots) that randomly fill out surveys. We filtered out bots by excluding annotations from an additional 107 annotators that marked more than $2/3$ of words as emphasized in eight or more utterances of the 20 utterances in a batch. Annotators who fail the bot filter are blocked from performing further annotation. We also recorded participants' native country and language, but note these may be unreliable as many MTurk workers use VPNs to subvert IP region filters on MTurk~\cite{moss_rosenzweig_jaffe_gautam_robinson_litman_2021}.

The average Cohen Kappa score for annotators with at least one overlapping utterance is 0.226 (i.e., ``Fair'' agreement)---but not all annotators annotate the same utterances, and this overemphasizes pairs of annotators with low overlap. Therefore, we use a one-parameter logistic model (i.e., a Rasch model) computed via \texttt{py-irt}~\cite{lalor2023py}, which predicts heldout annotations from scores of overlapping annotators with 77.7\% accuracy (50\% is random).


\section{Neural prominence estimation}
\label{sec:model}

\begin{figure}[t]
    \centering
    \includegraphics[width=\linewidth]{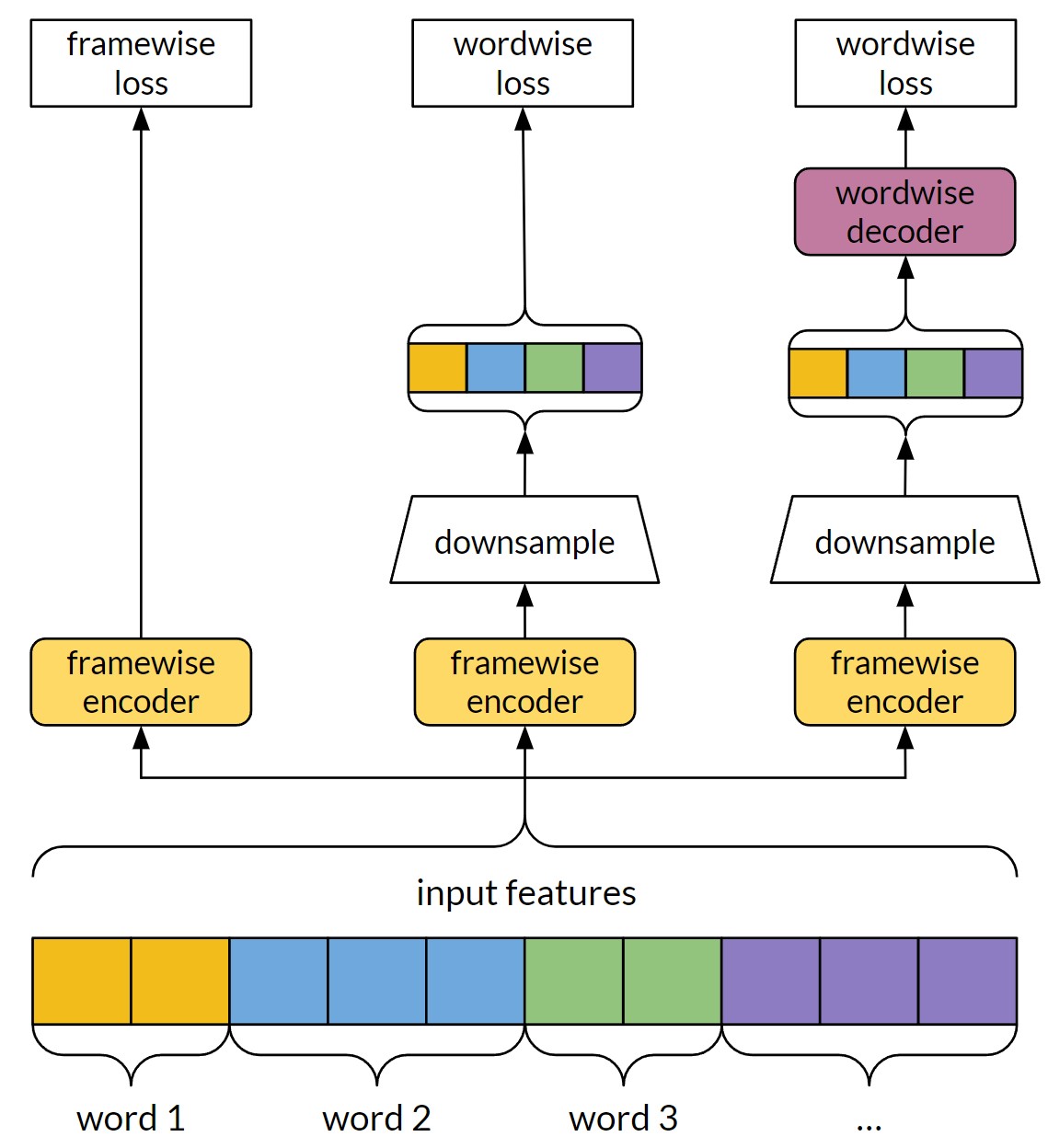}
    \caption{\textbf{Three candidate prominence estimation models $|$} We experiment with a \textbf{framewise} model \textbf{(left)} as well as two wordwise models: one that downsamples from frames to words just before the loss function (\textbf{posthoc wordwise}; \textbf{center}) and one that downsamples within the neural network (\textbf{intermediate wordwise}; \textbf{right}). The yellow, framewise encoder is a stack of convolution layers that operate at the frame resolution. The magenta, wordwise decoder is a stack of convolution layers that operate at the word resolution.}
    \label{fig:models}
    \vspace{-1em}
\end{figure}

We propose a neural network that predicts human-annotated prominence values for a sequence of words from acoustic features. Our proposed model takes as input an 80-channel Mel spectrogram) at an evenly-quantized time frame resolution (e.g., ten milliseconds), as well as a time-alignment between words and frames (i.e., the start and end frame indices corresponding to each word). We found no improvement when adding pitch, periodicity, or loudness features, which may be redundant with the spectrogram. Our network produces one prominence value per word. Crucial to such a system is the mechanism to perform variable-stride downsampling from the frame resolution to the word resolution. Vaidya et al.~\cite{9747780} propose one option (which we refer to as \textbf{prehoc wordwise}) that segments words before input to the model, such that the receptive field of the framewise encoder is limited to a single word. We identify three additional locations within the system for performing downsampling (Figure~\ref{fig:models}): (1) (\textbf{framewise}) upsample the ground truth prominence values to the frame resolution using linear interpolation during training and downsample the network output to word resolution during inference, (2) (\textbf{posthoc wordwise}) downsample the network output during training and inference, and (3) (\textbf{intermediate wordwise}) downsample to the word resolution within the network.  We further identify four methods for variable-stride downsampling from the frame resolution to the word resolution at each location: (1) (\textbf{average}) take the channel-wise average over all frames corresponding to a word, (2) (\textbf{max}) take the channel-wise maximum over all frames corresponding to a word, (3) (\textbf{sum}) take the channel-wise sum over all frames corresponding to a word, and (4) (\textbf{center}) take the value of the frame in the center of the word. Each framewise encoder and wordwise decoder is a stack of six convolution layers with 80 channels, a kernel size of three, and intermediate ReLU activation. Our hyperparameter search indicates six layers improves over five or seven layers; 80 channels improves over 64 or 128 channels; ReLU improves over leaky ReLU~\cite{maas2013rectifier}, GeLU~\cite{hendrycks2016gaussian}, or Swish~\cite{elfwing2018sigmoid}; convolution layers improve over Transformer~\cite{NIPS2017_3f5ee243} layers; and dropout does not improve performance when our model is not overparameterized.


\section{Evaluation}

We design our evaluation to determine whether our proposed neural prominence estimation models exceed the performance of previous automatic prominence estimation methods. We perform ablations to show the relative impact of design decisions, such as the method for downsampling from frames to words. Finally, we demonstrate the scaling behaviors of our best model as a function of two key cost factors: the number of emphasis-annotated utterances and the number of annotators per utterance.


\begin{table}[t]
\centering
\begin{tabular}{r|cccc}
\textbf{Downsampling location} & \multicolumn{4}{c}{\textbf{Downsampling Method}} \\
 & Average & Center & Max & \textbf{Sum}\\
\hline
Inference (framewise) & 0.102 & 0.153 & 0.102 & 0.137 \\
\textbf{Intermediate (wordwise)} & 0.656 & 0.438 & 0.674 & \textbf{0.675} \\
Posthoc (wordwise) & 0.440 & 0.385 & 0.623 & 0.645 \\
\hline
Prehoc \cite{9747780} (wordwise) & 0.670 & 0.471 & 0.670 & 0.656 \\
\end{tabular}
\caption{\textbf{Ablations of downsampling methods and locations $|$} Pearson correlations between estimated and ground truth prominence on the unseen Buckeye~\cite{pitt2005buckeye, cole2017crowd} dataset. Averages over three runs.}
\label{tab:downsample}
\vspace{-1em}
\end{table}

\subsection{Data}
\label{sec:data}

We train our models on our annotated LibriTTS partition (Section~\ref{sec:annotation}). We determine word boundaries by performing forced alignment between the speech transcript and audio. We use the Penn Phonetic Forced Aligner (P2FA)~\cite{scotuscorpus} via the Python Forced Alignment (\texttt{pyfoal}) library~\cite{Morrison_pyfoal_2021}. We extract from the 16 kHz audio 80 bands of a log-mel spectrogram using a hopsize of 160 samples and a window size of 1024 samples. We partition utterances into train (80\%), validation (10\%), and test (10\%) partitions. Scaling experiments use a different partitioning (see Section~\ref{sec:experiments}).

To examine generalization to unseen speakers, datasets, and speaking styles, we perform additional evaluation on emphasis annotations ~\cite{cole2017crowd} of the Buckeye corpus of conversational American English~\cite{pitt2005buckeye}. This evaluation dataset consists of 16 utterances from 16 speakers for a total of 256 utterances lasting 3.98 minutes and containing 931 words. Each utterance is annotated by 32 native speakers of American English.


\subsection{Training}
\label{sec:train}

We train using an Adam optimizer with a learning rate of $1e^{-3}$ to optimize a binary cross entropy (BCE) loss between predicted and ground truth prominence values. In Table~\ref{tab:ablate}, we show that mean squared error (MSE) and bounding the output to be between zero and one performs comparably with BCE, making MSE a viable option for loss function as well. We train for 6,000 steps, validating every 100 steps and checkpointing when the Pearson correlation between model output and human labeling (Section~\ref{sec:metrics}) is the maximum so far. We use a variable batch size~\cite{gonzalez2023batching} during training with a maximum of 75,000 frames (12.5 minutes) per batch.


\subsection{Metrics}
\label{sec:metrics}

We measure the accuracy of a prominence estimation system by calculating the Pearson correlation and BCE between the system's output and the corresponding ground truth prominence value for each word. BCE is equal to the KL divergence between Bernoulli random variables up to a dataset-dependent constant, making it well-suited for our probabilistic view of prominence as Bernoulli distribution parameters (Section~\ref{sec:intro}). We use Pearson correlation (instead of BCE) to compare with digital signal processing (DSP) baselines, as the range of system outputs varies between baseline systems and Pearson correlation is scale-invariant. 


\begin{table}[t]
\centering
\begin{tabular}{l|cc|cc}
& \multicolumn{2}{c|}{\textbf{Buckeye}} & \multicolumn{2}{c}{\textbf{LibriTTS}} \\
\textbf{Model} & PC$\uparrow$ & BCE$\downarrow$ & PC$\uparrow$ & BCE$\downarrow$\\
\hline
Proposed & \textbf{0.675} & \textbf{0.337} & \textbf{0.534} & \textbf{0.358} \\
\rowcolor{tablecolor}
\quad MSE loss & 0.672 & 0.340 & 0.533 & 0.361 \\
\hline
Wavelet~\cite{SUNI2017123, seshadri22_interspeech} & 0.529 & -- & 0.393 & -- \\
\rowcolor{tablecolor}
\end{tabular}
\caption{\textbf{Automatic prominence estimation results $|$} Pearson correlation (PC) and binary cross-entropy (BCE) between inferred and human annotations on two datasets. Includes our best model (the \textbf{intermediate wordwise} model with \textbf{sum} downsampling), one ablation, and a heuristic baseline. Averages over three runs.}
\label{tab:ablate}
\vspace{-1em}
\end{table}

\subsection{Experimental design}
\label{sec:experiments}

We now describe our experiments designed to demonstrate the performance of our model relative to previous works, the efficacy of our individual design choices, and the scaling behaviors of interest.

In Table~\ref{tab:downsample}, we experiment with all combinations of the four downsampling locations and four downsampling methods described in Section~\ref{sec:model}. In Table~\ref{tab:ablate}, we ablate our loss function and compare to a top-performing heuristic method (\textbf{wavelet})~\cite{seshadri22_interspeech} that classifies words by thresholding wavelet-based features proposed by Suni et. al~\cite{SUNI2017123}.

We demonstrate scaling behaviors for prominence estimation using our best model on curated data partitions of 400, 800, 1,600, and 3,200 utterances. We further analyze the cost efficiency of annotator redundancy by training on 400 annotations with eight annotators, 800 annotations with four annotators, 1,600 annotations with two annotators, and 3,200 annotations with one annotator. We discuss implications for cost-effective annotation.



\section{Results}
\label{sec:results}

The accuracies of our proposed neural prominence estimation method as well as baseline and ablation systems are reported in Tables~\ref{tab:downsample} and~\ref{tab:ablate}. In Table~\ref{tab:downsample}, we see that the variable-stride downsampling required for converting frame-resolution acoustic features to word-resolution prominence estimations is best performed within the network (\textbf{intermediate}) by taking the \textbf{sum} of each channel over the frames corresponding to each word. This outperforms the method proposed by Vaidya et al.~\cite{9747780} \textbf{(prehoc)} that segments the input acoustic features into words and indicates that it is efficacious for the receptive field of the framewise encoder to span across words. Relative to \textbf{prehoc}, our proposed \textbf{intermediate} method is also faster to train (by 39.7\%) and reduces GPU memory consumption (by 69.1\%) during training. In Table~\ref{tab:ablate}, we see that our best model significantly outperforms a top heuristic baseline (\textbf{wavelet}) in PC and BCE on both heldout data from our LibriTTS annotations and heldout data from an unseen dataset with speakers and speaking styles outside the training distribution (Buckeye).

\begin{figure}[t]
\centering
    \includegraphics[width=\linewidth]{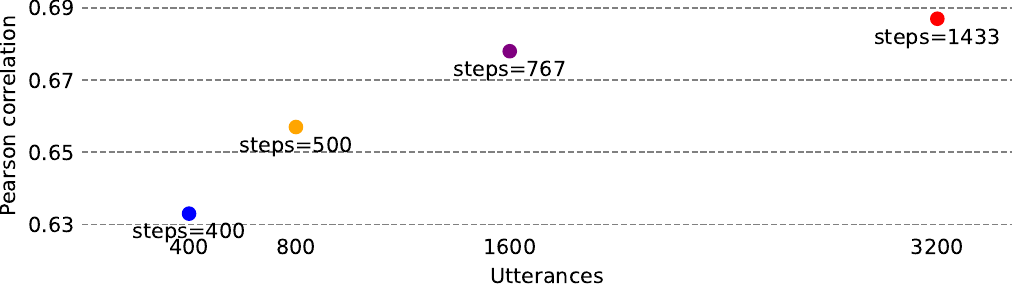}
    
    \vspace{1em}
    
    \includegraphics[width=\linewidth]{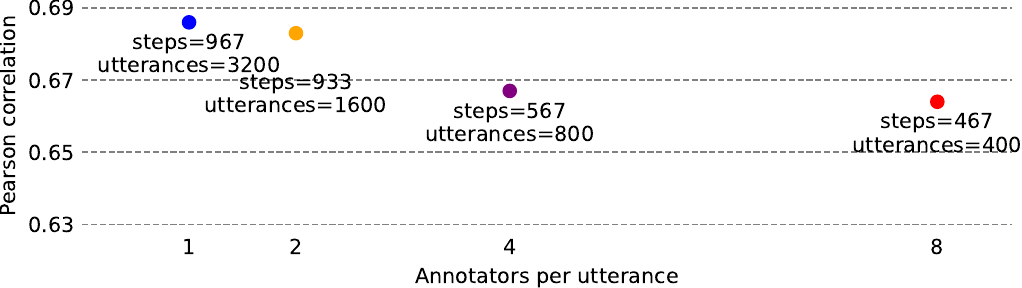}
\caption{\textbf{Scaling behaviors for neural prominence estimation $|$} \textbf{(top)} Pearson correlation between inferred and
human annotations as a function of dataset size. \textbf{(bottom)} Pearson correlation as a function of number of annotators for a fixed annotation budget. Averages over three runs.}
\label{fig:scaling}
\vspace{-1em}
\end{figure}

Figure~\ref{fig:scaling} demonstrates scaling behaviors for neural prominence estimation using our crowdsourced emphasis annotations of the LibriTTS dataset. We provide key takeaways:


\noindent
\textbf{For a fixed budget, use only one annotator per utterance $|$} Neural networks benefit from training on larger datasets. This benefit empirically outweighs the benefit of variance reduction in the ground truth distribution caused by annotator redundancy. However, when data is limited, annotator redundancy can improve performance. Cole et al.~\cite{cole2017crowd} note marginal variance reduction beyond seven annotators, so we do not expect these improvements to extrapolate to, e.g., 16 annotations per utterance.

\noindent
\textbf{For a fixed budget, increasing annotations per utterance up to eight improves convergence speed $|$} As the number of annotations per utterance increases from one to eight, the number of steps needed for convergence significantly decreases.

To further examine scaling behaviors, we used our best model to annotate the entire \texttt{train-clean-100} partition of LibriTTS and train a new model from scratch on 26,588 automatically annotated utterances. The resulting model performed marginally worse on heldout data, underscoring the utility of human annotations. 

\section{Conclusion}
\label{sec:conclusion}

The prominence of a spoken word is a fundamentally perceptual phenomenon. Our work highlights the benefits of utilizing human perception for prominence estimation, and demonstrates high-quality, generalizable prominence estimation trained from crowdsourced emphasis annotations \textbf{(Contribution 1)}. We further solve the lack of publicly available prominence annotations suitable for training a generalizable machine learning model \textbf{(Contribution 2)} and provide tools \textbf{(Contribution 3)} and guidelines \textbf{(Contribution 4)} for practitioners performing crowdsourced annotation. These contributions enable future work in high-quality emphasis-controlled text-to-speech, analysis of the human perception of prominence, and automatic detection and control of other word-level attributes (e.g., disfluency, falsetto, and vocal fry).


\bibliographystyle{IEEEbib}
\bibliography{refs}


\end{document}